\documentclass[prl,twocolumn,showpacs,preprintnumbers,amsmath,amssymb,twoside]{revtex4}

\usepackage{graphicx}
\usepackage{dcolumn}
\usepackage{bm}
\usepackage[pdftex, bookmarks={false}, pdfauthor={Rudro Rana Biswas}, pdftitle={Scattering from surface step edges in Strong Topological Insulators}]{hyperref}
\usepackage[all]{hypcap}

\newcommand\ba{\begin{array}}
\newcommand\ea{\end{array}}
\newcommand\bp{\begin{picture}}
\newcommand\ep{\end{picture}}
\newcommand\be{\begin{equation}}
\newcommand\ee{\end{equation}}
\newcommand\bs{\begin{subequations}}
\newcommand\es{\end{subequations}}
\newcommand\nn{\nonumber}
\newcommand\bfl{\begin{flushleft}}
\newcommand\efl{\end{flushleft}}
\newcommand\bsp{\begin{split}}
\newcommand\easp{\end{split}}

\newcommand\ri{\right}
\renewcommand\le{\left}

\newcommand{\sgn}{\mbox{sgn}}

\renewcommand\c{\psi}

\renewcommand\d{\delta}
\newcommand\D{\Delta}

\newcommand\G{\Gamma}

\newcommand\mbk{\mbs{k}}

\newcommand\p{\pi}

\newcommand\rr{\rho}
\newcommand\mbr{\mbs{r}}
\newcommand\s{\sigma}

\renewcommand\th{\theta}

\newcommand\Th{\Theta}

\newcommand\w{\omega}

\newcommand\imply{\Rightarrow}

\newcommand\la{\langle}
\newcommand\ra{\rangle}

\newcommand\mc{\mathcal}
\newcommand\mb{\mathbb}
\newcommand\mbs{\boldsymbol}

\begin{document}

\title{Scattering from Surface Step Edges in Strong Topological Insulators}
\author{Rudro R.\ Biswas$^{1,3}$}
\email{rrbiswas@physics.harvard.edu}
\author{Alexander V.\ Balatsky$^{2,3}$}%
\email{avb@lanl.gov}
\affiliation{%
$^{1}$Department of Physics, Harvard University, Cambridge, MA 02138\\
$^{2}$Theoretical Division, Los Alamos National Laboratory, Los Alamos, NM 87545\\
$^{3}$Center for Integrated Nanotechnologies, Los Alamos National Laboratory, Los Alamos, NM 87545
}
\date{\today}
\begin{abstract}
We study the characteristics of scattering processes at step edges on the surfaces of Strong Topological Insulators (STI), arising from restrictions imposed on the $S$-matrix \emph{solely} by time reversal symmetry and translational invariance along the step edge. We show that the `perfectly reflecting' step edge that may be defined with these restrictions allow modulations in the Local Density of States (LDOS) near the step edge to decay no slower than $1/x$, where $x$ is the distance from the step edge. This is faster than in 2D Electron Gases (2DEG) --- where the LDOS decays as $1/\sqrt{x}$ --- and shares the same cause as the suppression of backscattering in STI surface states. We also calculate the scattering at a delta function scattering potential and argue that \emph{generic} step edges will produce a $x^{-3/2}$ decay of LDOS oscillations. Experimental implications are also discussed.
\end{abstract}
\maketitle

Strong topological Insulators (STI) are three-dimensional band insulators that have an odd number of gapless chiral modes on their surfaces\cite{2007-fu-rt,2007-moore-yq,2009-zhang-vn}. These have recently been realized experimentally\cite{2009-xia-vn,2009-roushan-vn,2009-hsieh-fr} and are an active area of current research. The chiral states on the STI surfaces consist of time-reversed pairs of states propagating in opposite directions, between which backscattering by time-reversal invariant impurities and perturbations\cite{2009-roushan-vn} is forbidden. Because of this, we can expect scattering at impurities or step edges on the STI surface to lead to outcomes that are substantially different from the case of the 2D Electron Gas (2DEG). This suppression of backscattering is also often quoted as a major reason for the topological protection of the STI surface states against surface impurities. While this suppression is precise for the case of exact backscattering, it does not forbid scattering processes in which the particle gets reflected `almost' backward -- the formation of localized resonances/states near surface defects\cite{2010-yokoyama-fk,2009-biswas-rz} are still allowed. It is therefore instructive to test the response of these STI surface states to various surface defects and precisely quantify the restrictions imposed by the robust Time Reversal Symmetry (TRS). 
This knowledge will be an important ingredient in future attempts to design electronic devices based on these materials. Finally, the calculations in this paper also address a growing body of literature on the use of Scanning Tunneling Spectroscopy (STS) to investigate the role of isolated impurities and step edges on STI surfaces\cite{2010-alpichshev-fk,2009-roushan-vn,2009-gomes-pi}.

\begin{figure}[h]
\begin{center}
\resizebox{9cm}{!}{\includegraphics{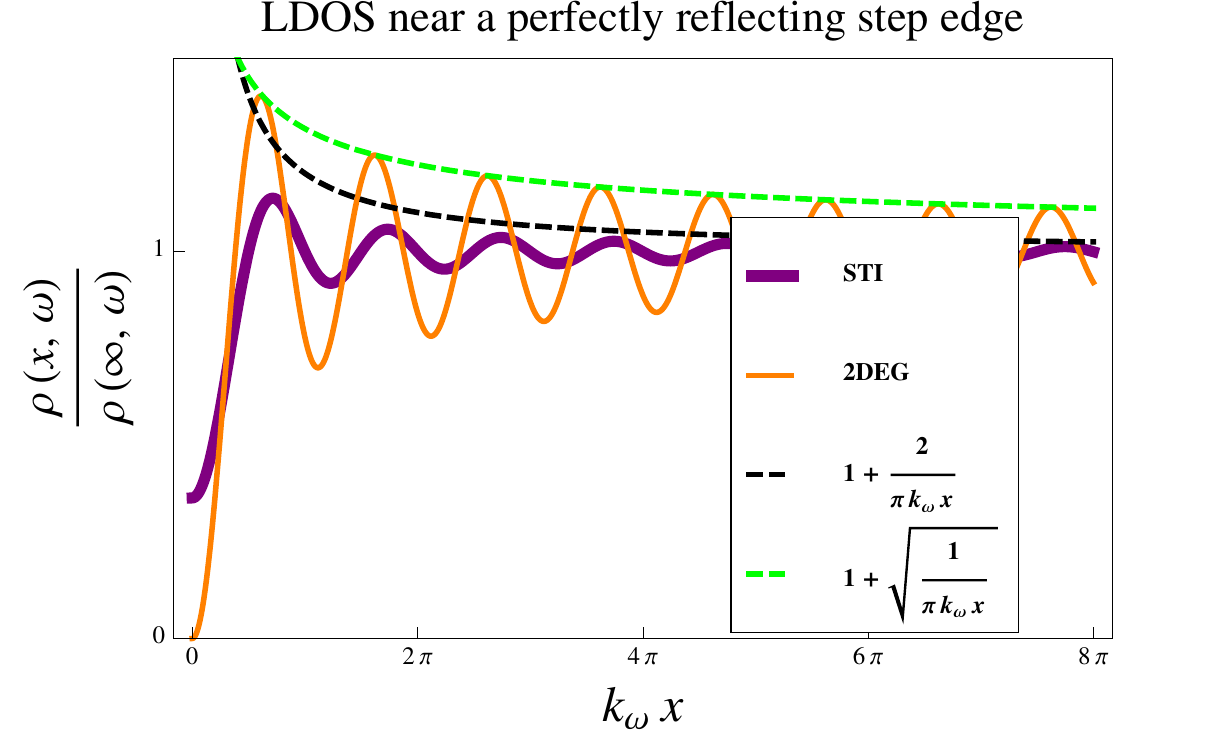}}
\caption{(Color Online) Comparison between the LDOS near a `perfectly reflecting' step edge in the STI \eqref{eq-perfectreflect} and the 2DEG. Their envelopes, decaying as $1/x$ and $1/\sqrt{x}$, respectively, are shown as dashed curves. $k_{\w}$ is the wave vector at energy $\w$.}
\label{fig-perfect}
\end{center}
\end{figure}

We shall address, in this paper, the scattering physics characteristic of a \emph{single} set of chiral states -- such as the Dirac states present on the surface of Bi$_{2}$Te$_{3}$ or Bi$_{2}$Se$_{3}$\cite{2009-hsieh-fr,2009-zhang-vn} --- in the context of scattering from a step edge. We find the following:\\
(i) TRS and unitarity impose a set of constraints on the reflection and transmission amplitudes (equation\ \eqref{eq-constraints1}) \emph{irrespective} of the effective Hamiltonian describing the step edge. We only require that the step edge does not violate TRS and is translationally invariant along its length.
\\
(ii) Suppressed backscattering leads to a substantial decrease of LDOS modulation near the step edge on the surface of the STI --- the LDOS is found to decay at least as fast as (Figure \ref{fig-perfect})
\begin{align}\label{eq-slowdecay}
\frac{\d\rr(\w,x)}{\rr^{(0)}(\w)} &\sim \frac{1}{x}
\end{align}
In contrast, in 2DEGs this decay is slower $\sim 1/\sqrt{x}$ \cite{1993-crommie-fk}.
\\
(iii) We predict the existence of the `perfectly reflecting' step edge \eqref{eq-perfectreflectance} by using the scattering matrix restrictions. This perfectly reflecting wall produces LDOS modulations of the kind mentioned in \eqref{eq-slowdecay} (shown in Figure \ref{fig-perfect}).
\\
(iv) For a sharp step edge, which we approximate using a delta function, we have evaluated the reflection amplitude explicitly as a function of the potential strength \eqref{eq-deltareflectance} and using this, find that for a `strong' potential there are essentially no oscillations near the step edge. The LDOS decays monotonically as $x^{-3/2}$ far from the edge, after an initial dip (Figure \ref{fig-delta}) -- similar to the long wavelength scattering observed in \cite{2010-alpichshev-fk}. The $x^{-3/2}$ decay law is also true for \emph{generic} step edges due to TRS.

We begin by formulating the general scattering matrix framework for the case of scattering at a surface step edge. We label the STI surface states by their band index $s = \pm 1$ and momentum $\mbk = (k_{x}, k_{y})$, which together also determine their energies $E_{s, \mbk}$. We choose $E=0$ at the Dirac point and $s = \sgn E_{s,\mbk}$. The defining characteristic of these states is that oppositely propagating states at a given energy are time-reversed partners ($\Th$ is the time reversal operator):
\begin{align}
\le|s,-\mbk\ri\ra \propto \Th\le|s, \mbk\ri\ra
\end{align}
\begin{figure}[h]
\begin{center}
\resizebox{8.5cm}{!}{\includegraphics{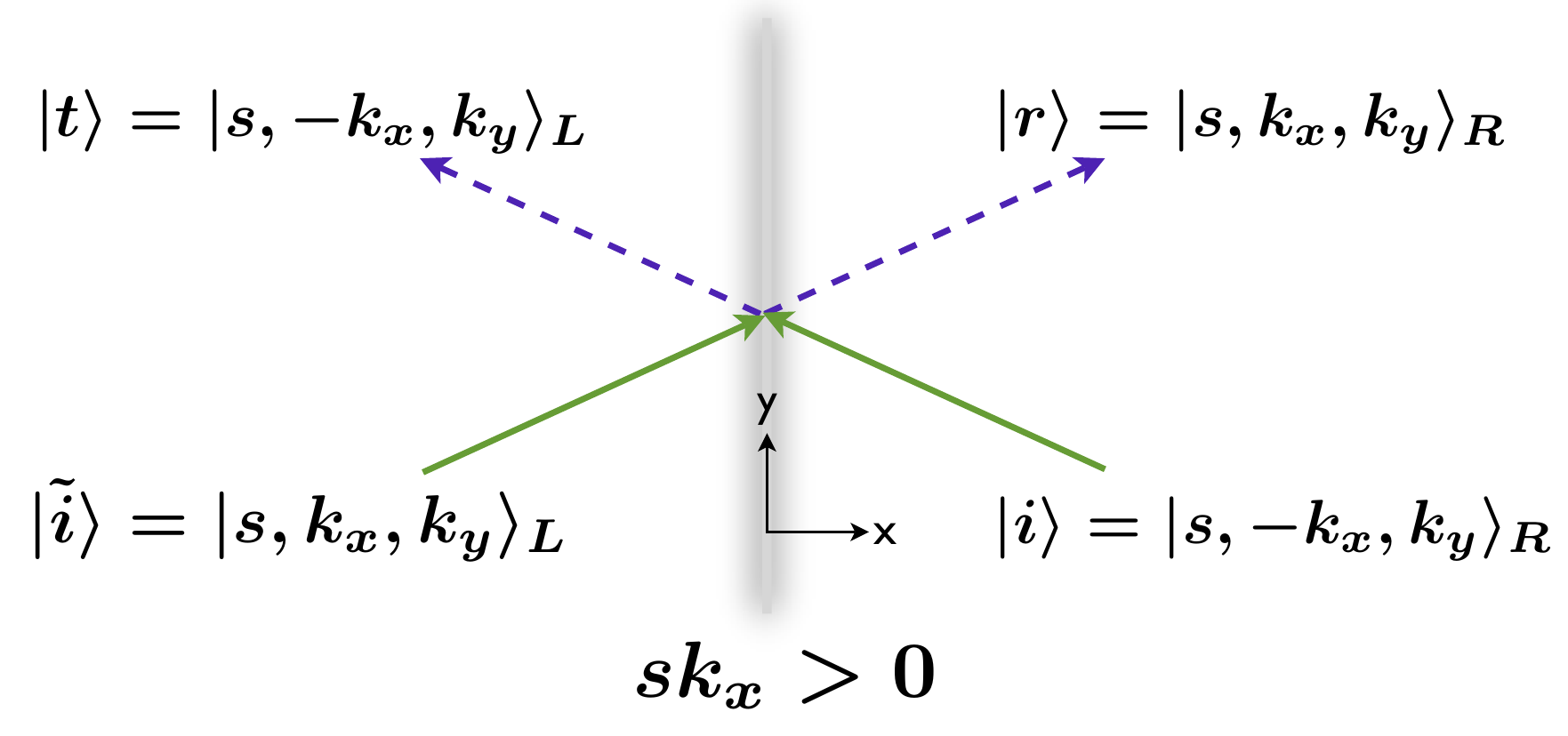}}
\caption{(Color Online) The scattering problem at a step edge, at an energy $E_{s, \mbk}$ and a $y$-momentum $k_{y}$. The green (incoming) and dashed violet (outgoing) arrows label the `incoming' and `outgoing' states respectively.}
\label{fig-scattering}
\end{center}
\end{figure}

We assume that there is translational invariance in the direction parallel to the step edge, say $y$ (See Figure \ref{fig-scattering}). Thus, the step edge can be represented by a finite-range modification to the Hamiltonian that preserves the $y$-momentum\cite{1991-davis-uq,1993-crommie-fk}. In some cases, it may be modeled as an electrostatic potential $V(x)$ that is a function of only the perpendicular coordinate $x$. Scattering processes at the step leave $k_{y}$ -- the $y$-momentum -- unchanged. In what follows, we shall first embody the effect of the step edge in a scattering matrix (the $S$-matrix). We label the scattering process as in Figure \ref{fig-scattering} and define the reflection and transmission coefficients to be ($r, r'$) and ($t, t'$) respectively. The energy of the scattering process is $E_{sk}$. The states are taken to belong to the $s^{\text{th}}$ band -- this is not mentioned explicitly, for brevity, in the following equations. $R/L$ denotes the half-space (right/left) these states belong to. The $S$-matrix is
\pagebreak
\begin{align}
&\mb{S}(k_x, k_y)= \le(\ba{cc} r(k_x, k_y) & t'(k_x, k_y) \\ t(k_x, k_y) & r'(k_x, k_y) \ea\ri)\\
&\equiv \le(\ba{cc} _{R}\la k_x, k_y|\hat{S}|-k_x, k_y\ra_{R} & _{R}\la k_x, k_y|\hat{S}| k_x, k_y\ra_{L} \\ _{L}\la -k_x, k_y|\hat{S}| -k_x, k_y\ra_{R} & _{L}\la -k_x, k_y|\hat{S}| k_x, k_y\ra_{L} \ea\ri)\nn
\end{align}

Step-edges don't usually break time-reversal symmetry unless, for e.g, they have embedded magnetic impurities. We can use this fact to find relations between the reflection/transmission amplitudes defined above. For simplicity, we fix the phases of the \emph{incoming} states by defining them to be the time-reversed versions of the appropriate outgoing states:
\begin{align}\label{eq-phases}
|-k_x, k_y\ra_{R} &\stackrel{s k_{x} > 0}{=} \Th |k_x, -k_y\ra_{R}\\
|k_x, k_y\ra_{L} &\stackrel{s k_{x} > 0}{=} \Th |-k_x, -k_y\ra_{L}\nn
\end{align}
Using this convention and the relation $\Th^{2} = -1$, we can prove the following constraints that are similar to the Stokes' relations in Optics:
\begin{align}\label{eq-constraints1}
r(k_x, k_y) = - r(k_x, -k_y)&,\; r'(k_x, k_y) = - r'(k_x, -k_y)\nn\\
t(k_x, k_y) &= - t'(k_x, -k_y)
\end{align}
As an example, we prove the first relation explicitly:
\begin{align}\label{eq-reflectance}
&r(k_x, k_y) \equiv _{R}\la k_x, k_y|\hat{S}|-k_x, k_y\ra_{R}\nn\\
&= _{R}\la \Th(-k_x, -k_y)|\hat{S}|-k_x, k_y\ra_{R}\nn\\
&=_{R}\la \Th(-k_x, k_y)|\hat{S}|\Th^{2}(-k_x, -k_y) \ra_{R}\nn\\
&= -_{R}\la k_x, -k_y|\hat{S}|-k_x, -k_y\ra_{R} =  - r(k_x, -k_y)
\end{align}
The antiunitary nature of the time reversal operator was used in the third line and $\Th^{2} = -1$, corresponding to spin-$1/2$ states, was used in the last step.

The unitary nature of the scattering process requires $\mb{SS}^{\dag} = \mb{I}$ and so:
\begin{align}\label{eq-unitarity}
|r|^{2}+|t|^{2}&=|r'|^{2}+|t'|^{2} = 1\nn\\
r^{*} t'+r' t^{*} &= 0\nn\\
(\imply |r|^{2}+|t'|^{2}&=|r'|^{2}+|t|^{2} = 1\mbox{ also})
\end{align}

If we are able to specify $\mb{S}$, the asymptotic forms of the new energy eigenstates, i.e, the wavefunctions outside the  influence of $V(x)$, are given by:
\begin{align}
|\mbk\ra_{R}^{\text{new}} &= \frac{|-k_x, k_y\ra_{R} + r(\mbk)\le|k_x, k_y\ri\ra_{R} + t(\mbk)\le|-k_x, k_y\ri\ra_{L}}{\sqrt{2}}\nn\\
|\mbk\ra_{L}^{\text{new}} &= \frac{|k_x, k_y\ra_{L} + r'(\mbk)\le|-k_x, k_y\ri\ra_{L} + t'(\mbk)\le|k_x, k_y\ri\ra_{R}}{\sqrt{2}}
\end{align}
The normalizations follow from \eqref{eq-unitarity}. Since we have these new eigenstates of the Hamiltonian, it is straightforward to calculate the modified LDOS away from the step edge:
\begin{widetext}
\begin{align}\label{eq-ldos1}
&\rr(\w,x>0) = \sum_{s=\pm1}\int_{0}^{s\cdot\infty}\frac{dk_{x}}{\p}\int_{-\infty}^{\infty}\frac{dk_{y}}{2\p}\le(|\c_{s,\mbk,R}^{\text{new}}(x,y)|^{2} + |\c_{s,\mbk,L}^{\text{new}}(x,y)|^{2}\ri)\d(\w - E_{s,\mbk})\nn\\
&= \rr^{(0)}(\w) + \iint_{0}^{\infty}\frac{dk_{x}dk_{y}}{\p^{2}}\text{Re}\le[r(s,sk_{x},k_{y})\c_{s,-sk_{x},k_{y}}^{(R)}(x,y)^{\dag}\c_{s,sk_{x},k_{y}}^{(R)}(x,y)\ri]\d(\w - E_{s,\mbk})_{s=\text{sgn}\w}
\end{align}
\end{widetext}
Here, $\rr^{(0)}$ is the LDOS in the absence of the step and simplifications have been made using time reversal symmetry. At this stage we can approximate the band structure near the Dirac point as that arising from the Dirac hamiltonian\footnote{The linear nature of the dispersion is not important -- the chiral nature of the Hamiltonian is.}
\begin{align}\label{eq-h0}
\mc{H} &= v \mbs{\s}\cdot\mbk
\end{align}
Here, $\mbs{\s}$ is the actual spin operator, up to a rotation. The \emph{outgoing eigenstates} of this Hamiltonian are given by ($sk_{x}\gtrless0$ for the $R/L$ cases):
\begin{align}
\c_{s,\mbk}^{(R/L)}(x,y) &= \frac{1}{\sqrt{2}}\le(\ba{c} e^{i\th_{\mbk}}\\ s \ea\ri) e^{i\mbk\cdot\mbr}, \; E_{s,\mbk} = s v k
\end{align}
The phases of the \emph{incoming eigenstates} are set by \eqref{eq-phases}:
\begin{align}
\c_{s,\mbk}^{(R/L)}(x,y) &= \frac{1}{\sqrt{2}}\le(\ba{c} s\\ e^{-i\th_{\mbk}} \ea\ri) e^{i\mbk\cdot\mbr}
\end{align}
We have used $\Th = i\s_{y}K$, $K$ being the complex conjugation operator, in the above equation. Using this setup, \eqref{eq-ldos1} can be simplified to yield (for $x>0$):
\begin{align}\label{eq-result1}
\frac{\d\rr(\w,x)}{\rr^{(0)}(\w)} &= - \frac{2\sgn\w}{\p}\!\!\int_{0}^{\p/2} \!\!\!\!\!\!\!\!d\th\,\sin\th\,\text{Im}\le[r_{\w}(\th)e^{2i k_{\w} x\cos\th}\ri]
\end{align}
Here, $\rr^{(0)}(\w) = |\w|/(2\p v^{2})$, $k_{\w} = \w/v$ is the momentum at energy $\w$ and the reflection amplitude $r_{\w}(\th_{\mbk})\equiv r(\sgn\w,\mbk_{\w})$ needs to obey the property \eqref{eq-reflectance}
\begin{align}\label{eq-result2}
r_{\w}(-\th) &= - r_{\w}(\th)
\end{align}

We now note that the `perfect' reflector needs to have the form (no transmittance except at normal incidence)
\begin{align}\label{eq-perfectreflectance}
r_{\w}^{\text{perf}}(\th) = e^{i\d_{\w,|\th|}}\sgn(\th)
\end{align}
which tells us that the LDOS becomes ($x>0$):
\begin{align}\label{eq-perfectreflect}
&\frac{\d\rr(\w,x)}{\rr^{(0)}(\w)} = - \frac{2\sgn\w}{\p}\int_{0}^{\p/2} \!\!\!\!\!\!d\th\,\sin\th\sin\le(2 k_{\w} x + \d_{\w,|\th|}\ri)\nn\\
&= - \frac{(\cos(2 k_{\w} x + \d_{\w}) - \cos\d_{\w})}{\sgn\w\,\p k_{\w} x} \;(\text{when $\d$ is $\th$-indep.})
\end{align}
In Figure \ref{fig-perfect}, these LDOS modulations are compared with those near a perfectly reflecting step edge in a 2DEG. We see that even for `perfect' reflection the step edge is only able to create LDOS oscillations that decay as $1/x$, as opposed to the slower $1/\sqrt{x}$ decay in the 2DEG case\cite{1993-crommie-fk}.

In a recent paper (Figures 3a-e in \cite{2010-alpichshev-fk}) oscillations of the form given by \eqref{eq-perfectreflect} have been observed at energies far from the Dirac point where the wavefunction wavelength is comparable to or shorter than the width of the step edge. However, in that region the band surface exhibits hexagonal warping\cite{2009-fu-fk} and the dominant scattering signature comes from the scatterings between the states at the adjacent $M$-points (the step edge is oriented perpendicular to the $\G M$ direction). These do \emph{not} involve the special processes near perfect backscattering in the STI that we have considered above. The $1/x$ behavior arises simply because the bands are not perpendicular to the $\G K$ direction near the hexagon vertices\footnote{For scattering between the hexagonal vertices, $r$ and the spin overlaps are finite and $\D k_{x}$ is a \emph{linear} function of $k_{y}$; a scaling analysis of \eqref{eq-ldos1} yields the $1/x$ law}.
\begin{figure}[h]
\begin{center}
\resizebox{9cm}{!}{\includegraphics{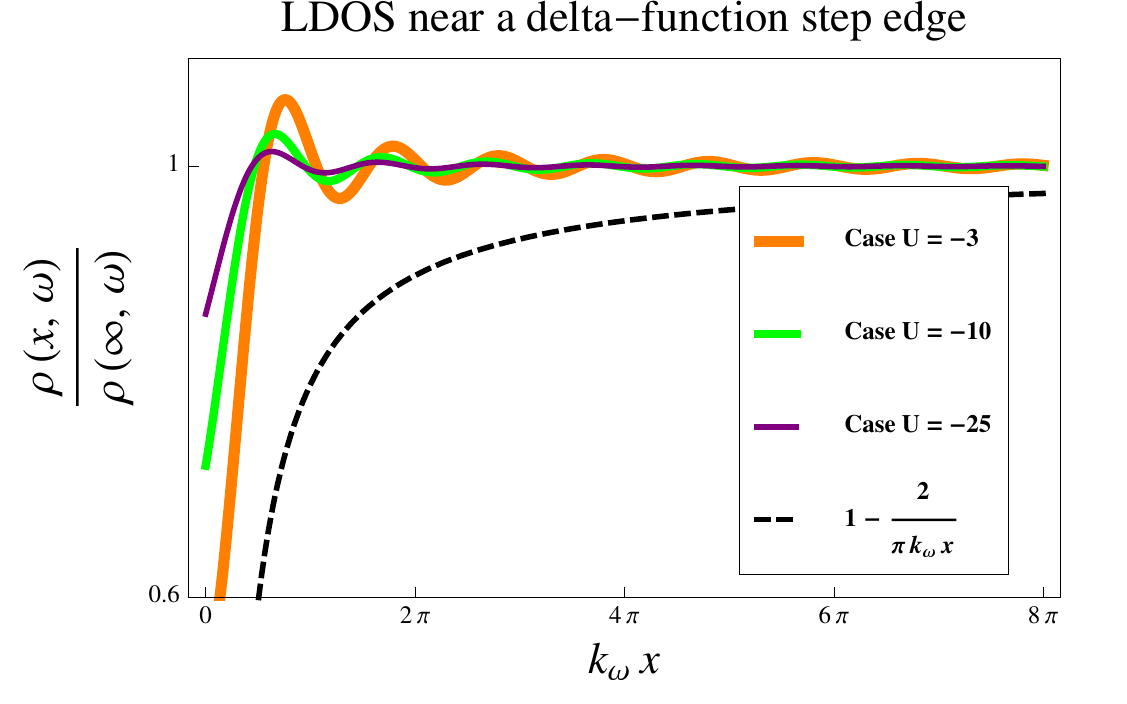}}
\caption{(Color Online) LDOS \emph{above} the Dirac point near a \emph{strong attractive} $\d$-function step edge potential \eqref{eq-Vdelta} for various values of $U$ (solid curves with different colors and thicknesses). Notice how the oscillations get damped for large $U$. The dashed curve is the envelope for the `perfectly reflecting' step edge on the STI (Figure \ref{fig-perfect}), shown here for comparison.}
\label{fig-delta}
\end{center}
\end{figure}
In Figures 3f-h of the above-mentioned paper \cite{2010-alpichshev-fk} the authors also observed that at lower energies, closer to the Dirac point, the character of these LDOS modulations change qualitatively. They are much less pronounced exhibit an almost oscillation-less decay. We propose to explain this behavior by modeling the step edge as a delta-function potential
\begin{align}\label{eq-Vdelta}
V(x) &= U \d(x)
\end{align}
when considering the scattering of long wavelength states near the Dirac point. Using this with the original Hamiltonian \eqref{eq-h0}, we obtain the reflection amplitude\cite{2006-katsnelson-kx}:
\begin{align}\label{eq-deltareflectance}
r_{\w}^{\d}(\th) = \frac{4 U \sin (\theta )}{\left(U^2-4\right) \cos (\theta )-4 i \sgn\w U}
\end{align}
Note that the reflection amplitude is antisymmetric about $\th = 0$, and is also linear around that region. The LDOS obtained for this delta-function potential is shown in Figure \ref{fig-delta} and displays minimal oscillations when $U$ is large and \emph{negative} and $E>0$. There is a dip in the LDOS near the wall followed by an almost oscillation-less decay (for large potential strength) away from the wall -- similar to what was observed in \cite{2010-alpichshev-fk}. The dip can be explained by the existence of \emph{bound states} near the wall: for $U<-2$ these occur at negative energies but the transferred spectral weight leads to the dip at positive energies.

The long-distance decay of the oscillations is given by $x^{-3/2}$, which arises out of the proportionality between $r(\th)$ and $\th$ near $\th=0$. Because of the antisymmetry of $r(\th)$ about $\th=0$ \eqref{eq-result2}, for \emph{generic} step edges we will also have $r(\th)\propto \th$ near $\th=0$ and this will result in the almost coherent scatterings near $\th=0$ interfering far away from the edge to yield a spatial decay law of $x^{-3/2}$\cite{2009-zhou-fk}. This power law may be obtained by considering the scaling behavior of the integrand in \eqref{eq-result1}, near $\th=0$.

It is possible that for high energies, near a step edge perpendicular to the $\G M$ direction in Bi$_{2}$Te$_{3}$ (or with \emph{any} orientation in Bi$_{2}$Se$_{3}$ which has no hexagonal warping), one will observe a $1/x$ decay of oscillations at the wave-vector equal to the diameter of the band surface along the $K-\G-K$ direction \emph{if} the step edge is perfectly non-reflecting -- a quality that possibly arises due to roughness along its normal direction. Generically, due to the scaling relation $r(\th)\propto \th$ near $\th=0$, one should be able to observe a faster $x^{-3/2}$ decay of the oscillations.

In summary, we have used the \emph{chiral nature} of the STI surface states and \emph{time reversal invariance} to impose restrictions on the reflection and transmission amplitudes of the scattering process at a step edge on the surface of a Strong Topological Insulator. This allowed us to define the `perfect reflecting step edge' for these chiral surface states and we found that the amplitude of LDOS ripples caused by such an edge decay no slower than $1/x$ --- faster than in the analogous case for a 2DEG, where the decay occurs as $1/\sqrt{x}$. For a \emph{generic} smooth step edge, the decay law is $x^{-3/2}$. We give possible reasons for the LDOS features seen in experiments\cite{2010-alpichshev-fk} --- damped oscillations and a monotonic decay of the LDOS modulation far from the step edge for small energies and a $1/x$ decay at higher energies -- by arguing that the scattering of long wavelength states near the Dirac point occurs due to an effective $\d$-function local potential \eqref{eq-Vdelta} while the shape of the band surface yields the $1/x$ decay at higher energies. To the best of our knowledge, this is the first application of this general $S$-matrix formalism (with the imposition of TRS) to the scattering of STI surface states.

We are grateful to D.\ Basov, Z.\ Hasan, S.\ Iyer-Biswas, H.\ Manoharan, N.\ Nagaosa, P.\ Roushan, H.\ Beidenkopf, Y.\ Xia and A.\ Yazdani for useful discussions. We would like to especially thank the Yazdani group for pointing out that the step edge used in \cite{2010-alpichshev-fk} was oriented perpendicular to the $\G M$ direction and that the observed oscillations occurred at the nesting wave vector, and N.\ Nagaosa for telling us about possible bound states in the system. This work was supported by the US DOE thorough BES and LDRD and by the University of California UCOP program T027-09.

\end{document}